\documentclass[graybox]{svmult}

\usepackage{mathptmx}       % selects Times Roman as basic font
\usepackage{helvet}         % selects Helvetica as sans-serif font
\usepackage{courier}        % selects Courier as typewriter font
\usepackage{type1cm}        % activate if the above 3 fonts are
\usepackage{makeidx}         % allows index generation
\usepackage{graphicx}        % standard LaTeX graphics tool
\usepackage{multicol}        % used for the two-column index
\usepackage[bottom]{footmisc}% places footnotes at page bottom

\makeindex             % used for the subject index
\usepackage{braket}
\usepackage{url}

\begin{document}

\title*{Quantum Diffusion in Separable d-Dimensional Quasiperiodic Tilings}
\author{Stefanie Thiem and Michael Schreiber}
\institute{Stefanie Thiem \at Institut f\"ur Physik, Technische Universit\"at Chemnitz, D-09107 Chemnitz, Germany \email{stefanie.thiem@physik.tu-chemnitz.de}
\and Michael Schreiber \at Institut f\"ur Physik, Technische Universit\"at Chemnitz, D-09107 Chemnitz, Germany \email{schreiber@physik.tu-chemnitz.de}}
\maketitle

\abstract{We study the electronic transport in quasiperiodic separable tight-binding models in one, two, and three dimensions. First, we investigate a  one-dimensional quasiperiodic chain, in which the atoms are coupled by weak and strong bonds aligned according to the Fibonacci chain. The associated $d$-dimensional quasiperiodic tilings are constructed from the product of $d$ such chains, which yields either the square/cubic Fibonacci tiling or the labyrinth tiling. We study the scaling behavior of the mean square displacement and the return probability of wave packets with respect to time. We also discuss results of renormalization group approaches and lower bounds for the scaling exponent of the width of the wave packet.}

%------------------------------------------------------------------------------------------------------------------------------------------
\section{Intoduction}
\label{sec:introduction}
%------------------------------------------------------------------------------------------------------------------------------------------

Understanding the relations between the atomic structure and the physical properties of materials remains one of the elementary questions of condensed-matter physics further emphasized by the discovery of quasicrystals \cite{PhysRevLett.1984.Shechtman}. Quasicrystals are characterized by a perfect long range order without having a three-dimensional translational periodicity. The former is manifested by the occurrence of sharp spots in the diffraction pattern and the latter in the occurrence of rotational symmetries forbidden for conventional crystals. Already in the 1970s, works by Penrose and Ammann showed that the Euclidean space can be filled gapless and non-overlapping by two or more tiles which are arranged in a
nonperiodic way according to matching rules. It turned out that these tilings are suitable to describe the structure of quasicrystals.

Experimental studies revealed rather exotic physical properties. For instance, quasicrystalline surfaces are anti-adhesive in combination with a high level of hardness making them suitable for the production of coatings for medical equipment, engines, cookware, etc. Further, they possess a low thermal and electrical conductance although they contain a high amount of well-conducting elements. For one-dimensional quasicrystals many numerical studies helped to establish a better understanding of the physical properties \cite{PhysRevB.1987.Kohmoto, JPhys.1995.Zhong, PhysRevB.2004.Sanchez}. Further, several exact theoretical results are known in one dimension today \cite{ComMathPhys.2011.Damanik}. However, the characteristics in two or three dimensions have been clarified to a much lesser degree because numerical studies are usually restricted to small approximants.

To address this problem we study models of $d$-dimensional quasicrystals with a separable Hamiltonian in a tight-binding approach \cite{EurophysLett.1989.Sire}. This method is based on the Fibonacci sequence, which describes the weak and strong couplings of atoms in a quasiperiodic chain.
After $a$ iterations of the inflation rule $\mathcal{P} = \{ s \longrightarrow w, w \longrightarrow ws \}$, we obtain the $a$th order approximant $\mathcal{C}_a$ of the Fibonacci chain. The length $f_a$ of an approximant $\mathcal{C}_a$ is given by the recursive rule $f_a = f_{a-1} + f_{a-2}$ with $f_0 = f_1 = 1$. Further, the ratio of the lengths of two successive approximants approaches the golden mean $\tau$ for $a \to \infty$.
Solving the time-independent Schr\"odinger equation
\begin{equation}
 \label{equ:octonacci.8}
 \mathbf{H} \ket{\Psi^i}  =  E^i \ket{\Psi^i} \Leftrightarrow t_{l-1,l} \Psi_{l-1}^i + t_{l,l+1} \Psi_{l+1}^{i} =  E^i \Psi_l^i
 \end{equation}
for the Fibonacci chain, we obtain discrete energy values $E^i$ and wave functions $\ket{\Psi^i} = \sum_{l=1}^{f_a+1} \Psi_l^i \ket{l}$ represented in the orthogonal basis states $\ket{l}$ associated to a vertex $l$.
The hopping strength $t$ in the Schr\"odinger equation is given by the Fibonacci sequence $\mathcal{C}_a$ with $t_{s} = s$ for a strong bond and $t_{w} = w$ for a weak bond ($0 < w \le s$).

\begin{figure}[t!]
 \sidecaption
 \includegraphics[width=3.3cm]{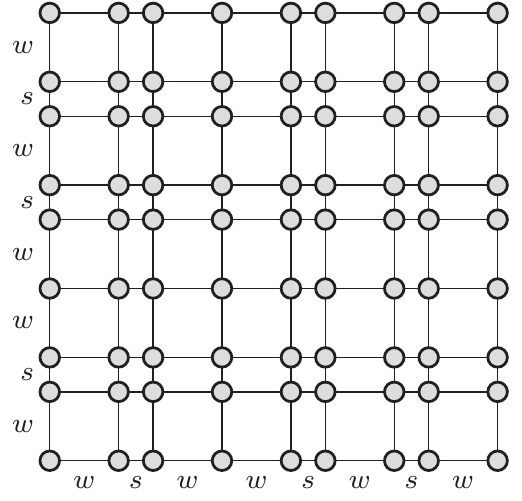}
 \includegraphics[width=3.3cm]{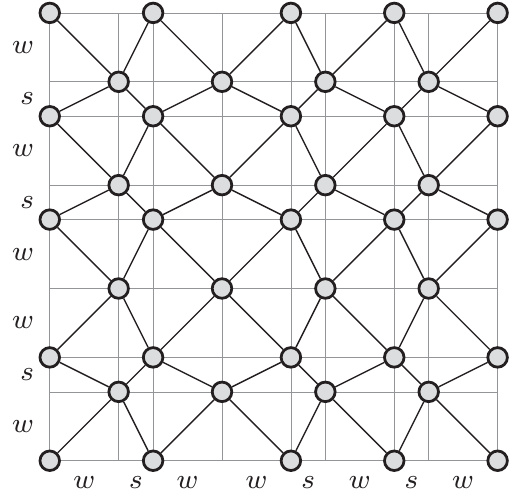}\vspace{-0.28cm}
 \caption{Two-dimensional square tiling $\mathcal{H}_5^{2{\rm d}}$ (left) and labyrinth tiling $\mathcal{L}_5^{2{\rm d}}$ (right) constructed from two Fibonacci chains $\mathcal{C}_5$.}
  \label{fig:tilings}
\end{figure}

The $d$-dimensional separable quasiperiodic tilings are then constructed from the product of $d$ quasiperiodic chains which are perpendicular to each other. For this setup two special cases are known for which the systems becomes separable \cite{EurophysLett.1989.Sire}:
\begin{itemize}
 \item \emph{Hypercubic Tiling $\mathcal{H}_a^{d{\rm d}}$ } --- This tiling corresponds to the usual Euclidian product of $d$ linear quasiperiodic chains, i.e., only vertices connected by vertical and horizontal bonds interact as shown in Figure \ref{fig:tilings}. The electronic structure and the transport properties are understood quite well for these systems \cite{JPhys.1995.Zhong,PhilMag.2008.Mandel}.
 \item \emph{Labyrinth Tiling $\mathcal{L}_a^{d{\rm d}}$} --- Only coupling terms to neighbors along the diagonal bonds are considered (cf.\ Figure \ref{fig:tilings}), where the bond strengths of this tiling equal the products of the corresponding bond strengths of the one-dimensional chains.
\end{itemize}
The eigenstates of these tilings in $d$ dimensions can be constructed from the eigenstates of $d$ one-dimensional chains. In both cases the wave functions of the higher-dimensional tiling are constructed as the products of the one-dimensional wave functions, i.e.,
   $ \Phi_\mathbf{r}^\mathbf{s} = \Phi_{l,m,\ldots, n}^{i,j,\ldots, k} \propto \Psi_{l}^{1i} \Psi_{m}^{2j} \ldots \Psi_{n}^{dk}$.
The superscripts $\mathbf{s} = (i,j,\ldots,k)$ enumerate the eigenvalues $E$, and $\mathbf{r}=(l,m,\ldots,n)$ denotes the coordinates of the vertices in the tiling. The energies are treated in a different way. In $d$ dimensions they are given by  $  E^\mathbf{s} = E^{i,j,\ldots ,k} = E^{1i} + E^{2j} + \ldots +  E^{dk} $ for the hypercubic tiling and by
$   E^\mathbf{s} = E^{i,j,\ldots ,k} = E^{1i} E^{2j}\ldots  E^{dk} $ for the labyrinth tiling. This approach allows us to study very large systems in higher dimensions with up to $10^{10}$ sites.

%------------------------------------------------------------------------------------------------------------------------------------------
\section{Quantum Diffusion}
%------------------------------------------------------------------------------------------------------------------------------------------

To obtain a deeper understanding of the connections of the electronic transport and the quasiperiodic structure of a system, we study the time evolution of wave packets $\ket{\Upsilon(\mathbf{r}_0,t)} = \sum_{\mathbf{r} \in \mathcal{L}} \Upsilon_\mathbf{r}(\mathbf{r}_0,t) \ket{\mathbf{r}}$ initially localized at a position $\mathbf{r}_0$ which are constructed from the solutions $ \Upsilon_{\mathbf{r}}(\mathbf{r}_0,t)
  = \sum_{\mathbf{s}} \Phi_{\mathbf{r}_0}^{\mathbf{s}} \Phi_{\mathbf{r}}^{\mathbf{s}}  e^{-\mathrm{i} E^{\mathbf{s}} t}$ of the time-dependent Schr\"odinger equation. The temporal autocorrelation function  of a wave packet equals the integrated probability to be at the position $\mathbf{r}_0$ up to time $t > 0$ \cite{PhysRevLett.1992.Ketzmerick}, i.e., $C(\mathbf{r}_0,t) = \frac{1}{t} \int_0^t |\Upsilon_{\mathbf{r}_0}(\mathbf{r}_0,t^\prime)|^2 \mathrm{d} t^{\prime}$.
We denote the integrand as return probability
 \begin{equation}
 \label{equ:return-prob}
   P(\mathbf{r}_0,t)  = |\Upsilon_{\mathbf{r}_0}(\mathbf{r}_0,t)|^2 \;.
 \end{equation}
 Another quantity often considered for the description of the electronic transport properties is the mean square displacement of the wave packet
(also called width)
\begin{equation}
 d(\mathbf{r}_0,t) = \sqrt{\sum\nolimits_{\mathbf{r} \in \mathcal{L}} |\mathbf{r}-\mathbf{r}_0|^2 \, |\Upsilon_\mathbf{r}(\mathbf{r}_0,t)|^2 } \;.
\end{equation}
The wave-packet dynamics reveal anomalous diffusion with $d(\mathbf{r}_0, t) \propto t^{\beta(\mathbf{r}_0)}$, and the electronic transport is governed by the wave-packet dynamics averaged over different initial positions $\mathbf{r}_0$, i.e., $d(t) = \langle d(\mathbf{r}_0,t) \rangle \propto t^{\beta}$. The exponent $\beta$ is related to the conductivity $\sigma$ via the generalized Drude formula, where $\beta=0$ corresponds to no diffusion, $\beta=1/2$ to classical diffusion, and $\beta=1$ to ballistic spreading.

\begin{figure}[t!]
 \centering
 \includegraphics[width=0.328\textwidth]{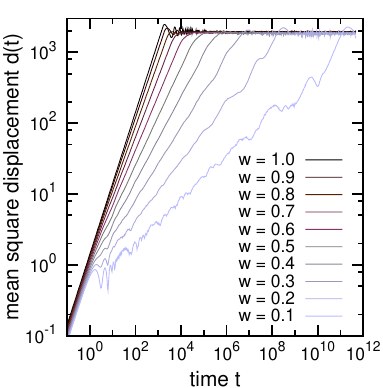}
 \includegraphics[width=0.328\textwidth]{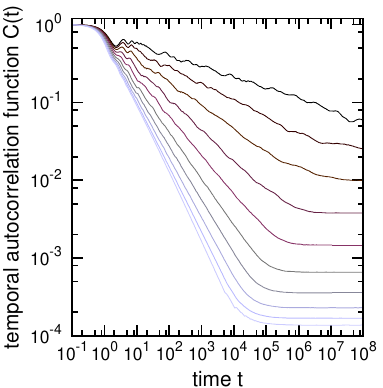}
 \includegraphics[width=0.328\textwidth]{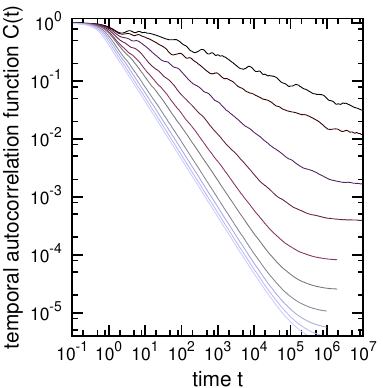}\vspace{-0.18cm}
 \caption{Averaged mean square displacement $d(t)$ for the Fibonacci chain $\mathcal{C}_{19}$ (left), averaged temporal autocorrelation function $C(t)$ for $\mathcal{C}_{20}$ (center) and the 2D labyrinth tiling $\mathcal{L}_{16}^{2{\rm d}}$ (right), $s = 1$.}
 \label{fig:autocorr}
\end{figure}

\begin{figure}[b!]
  \includegraphics[width=0.33\textwidth]{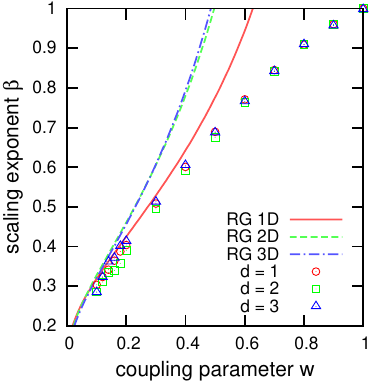}
  \includegraphics[width=0.33\textwidth]{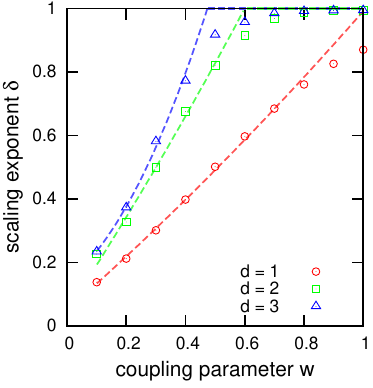}
  \includegraphics[width=0.33\textwidth]{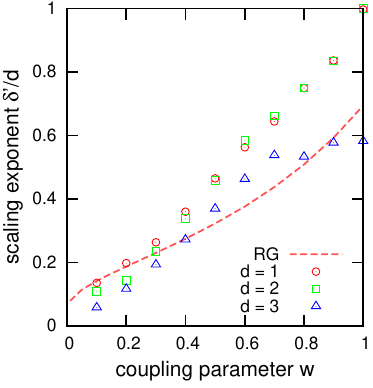}\vspace{-0.18cm}
  \caption{Scaling exponents $\beta$ (left), $\delta$ including the limit behavior (lines) for infinite systems (center), and $\delta^\prime/d$ (right) for the Fibonacci chain and the labyrinth tilings in 2D and 3D for $s=1$. RG results are shown by lines in left and right panel.}
  \label{fig:delta}
\end{figure}

The autocorrelation function is expected to decay with $C(\mathbf{r}_0,t) \propto t^{-\delta(\mathbf{r}_0)}$ \cite{PhysRevLett.1992.Ketzmerick}. The exponent $\delta(\mathbf{r}_0)$ is equivalent to the correlation dimension $D_2^\mu$ of the local density of states (LDOS) $\varrho(\mathbf{r}_0,E)$ of a system \cite{PhysRevLett.1992.Ketzmerick}.
In one dimension $\delta \rightarrow 1$ corresponds to the ballistic motion of an electron.
More information about the transport can be obtained by studying the return probability which shows a power-law behavior according to $P(t) \propto t^{-\delta^\prime}$. The integration is only used to smooth the results. However, this leads to some disadvantages: For $\delta^\prime = 1$ one obtains an additional logarithmic contribution $C(t) \propto \ln(t) / t$ leading to $\delta$ significantly lower than $1$ \cite{JPhys.1995.Zhong}. Further, in higher dimensions $\delta^\prime > 1 $ is possible  for large coupling parameters $w$. This leads to the convergence of the integral in $C(t)$ as shown in Figures \ref{fig:autocorr} and \ref{fig:delta} \cite{JPhys.1995.Zhong}.

Some typical results for the mean square displacement $d(t)$ and the temporal autocorrelation function $C(t)$ are shown in Figure \ref{fig:autocorr}. We average the results over different initial positions of the wave packet, i.e., we study the scaling behavior of $d(t) = \langle d(\mathbf{r}_0,t) \rangle \propto t^{-\beta}$ and $C(t) = \langle C(\mathbf{r}_0,t) \rangle \propto t^{-\delta}$.
The power-law behavior can be observed over several orders of magnitude in time, before $d(t)$ and $C(t)$ approach a constant due to finite size effects. The corresponding scaling exponents $\beta$ and $\delta$ obtained by least squares fits are compiled in Figure \ref{fig:delta}.
We find $\delta^\prime > 1$ for $w > 0.55 s$ in two dimensions and $w > 0.43 s$ in three dimensions. Consequently, the scaling exponent $\delta$ slowly approaches 1 for this threshold $w$ as indicated by the dashed lines \cite{PhysRevB.2005.Cerovski}. This is in reasonable agreement with the known transition to an absolute continuous energy spectrum for $w_{\rm th}^{\rm 2d} \approx 0.6 s$ and $w_{\rm th}^{\rm 3d} \approx 0.46 s$ \cite{PhilMag.2008.Mandel,PhysRevB.2000.Yuan}.
The exponent $\beta$ indicates anomalous transport for the hypercubic and the labyrinth tiling for all $w < s$. Only for $w=s$ ballistic transport is found.
Further, we observe that the exponent $\delta_{\rm 2d}^\prime$ is about twice as large as the one-dimensional exponent $\delta_{\rm 1d}^\prime$. However, the results for the three-dimensional labyrinth tiling do not entirely fit into this scheme. While for $w \le 0.6 s$ the expression $\delta_{\rm 3d}^\prime/d$ is only slightly smaller than the exponent $\delta_{\rm 1d}^\prime$, the behavior changes completely for large values of $w$. In the latter case the scaling exponent $\delta_{\rm 3d}^\prime$ becomes almost constant and approaches $\delta_{\rm 3d}^\prime(w=s) \approx 1.75$.

For the hypercubic tiling the relation $ \delta_{\rm 1d}^\prime = \delta_{d{\rm d}}^\prime/d$ holds, which also characterizes the transition to an absolutely continuous energy spectrum according to $d \delta_{\rm 1d} (w_{\rm th}^{d{\rm d}}) = 1$ \cite{JPhys.1995.Zhong}. While the numerical results for the 2D labyrinth tiling seem to satisfy these relations as well, there are significant differences in the wave-packet dynamics in three dimensions, which are probably related to the different grid structure. In three dimensions for $w \to s$ the labyrinth tiling approaches a \emph{body centered cubic} lattice while the cubic tiling approaches a \emph{simple cubic} lattice.

%------------------------------------------------------------------------------------------------------------------------------------------
\section{RG Approach and Lower Bound for the Scaling Exponent $\beta$}
\label{subsec:lower-bound}
%------------------------------------------------------------------------------------------------------------------------------------------

For the hypercubic tiling the scaling exponents of the width fulfill the relation $ \beta_{d{\rm d}} = \beta_{\rm 1d} $ due to the separability of the time-evolution operator. Numerical results suggest that this relation is also valid for the labyrinth. For weak couplings ($w \ll s$) it is possible to find a connection between the quasiperiodic structure of the Fibonacci chain \cite{PhysRevA.1987.Abe} or the labyrinth tiling \cite{PhysRevB.2012.Thiem,Thesis.2012.Thiem} with the wave-packet dynamics by an RG approach proposed by Niu and Nori \cite{PhysRevLett.1986.Niu}. The results show that the scaling exponents $\beta$ for different dimensions approach each other for $w \to 0$ but are not identical \cite{PhysRevB.2012.Thiem,Thesis.2012.Thiem} (cf.\ Figure \ref{fig:delta}).
The hierarchic structure of the RG enables us to describe the properties of the wave functions and energy spectrum of the Fibonacci chain for $w \ll s$  \cite{PhysRevLett.1986.Niu, PhysRevLett.1996.Piechon}. This can be used to derive analytical relations, too, for the mean-square displacement \cite{PhysRevLett.1996.Piechon} and the return probability (cf.\ Figure \ref{fig:delta}) \cite{Thesis.2012.Thiem,JPhysCondMatt.2012.Thiem}.

\begin{figure}[b!]
  \centering
  \includegraphics[width=0.328\textwidth]{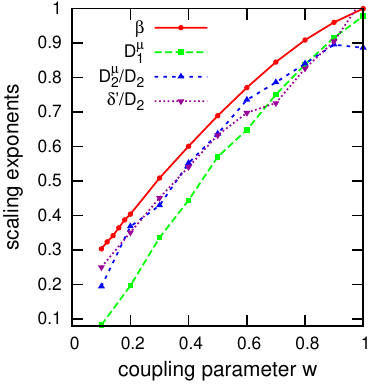}
  \includegraphics[width=0.328\textwidth]{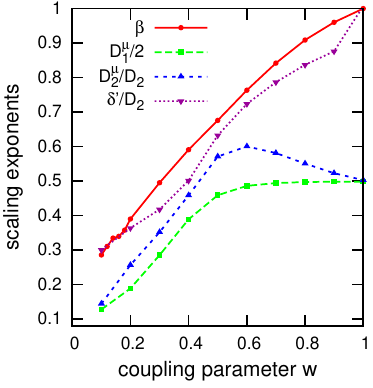}
  \includegraphics[width=0.328\textwidth]{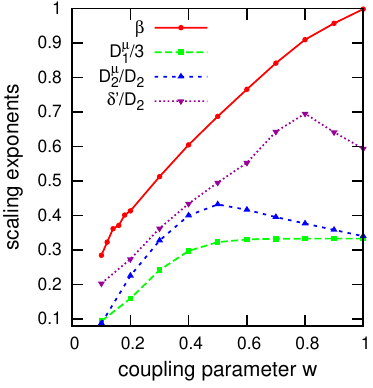}\vspace{-0.18cm}
  \caption{Comparison of $\beta$ with the lower bounds $D_1^\mu / d$, $D_2^\mu / D_2$, and $\delta^\prime / D_2$ for the Fibonacci chain (left) and the corresponding labyrinth tilings in two (center) and three dimensions (right) for $s =1$.}
  \label{fig:bounds-beta}
\end{figure}

The computation of the scaling exponent $\beta$ is computationally very expensive. However, one can make use of two general lower bounds for $\beta$. As a rule of thumb, the wave-packet propagation is faster for smoother spectral measures. This was proved by Guarneri et al. with the lower bound $\beta \ge D_1^\mu / d$ based on the information dimension $D_1^\mu$ of the spectral measure of the LDOS \cite{EurophysLett.1993.Guarneri}. Ketzmerick et al. showed that the spreading of a wave packet in a space of reduced dimension $D_2$ (i.e.\ the correlation dimension of the wave functions) is described by the bound $\beta \ge D_2^\mu/D_2$ as long as $D_2^\mu < 1$, i.e., for singular continuous energy spectra \cite{PhysRevLett.1997.Ketzmerick}. This result is based on the normalization condition and the known decay of the center of the wave packet according to $t^{-D_2^\mu}$.
A comparison with the numerical results in Figure \ref{fig:bounds-beta} shows that all inequalities are clearly satisfied.
However, in the regime of an absolutely continuous energy spectrum, the spectral dimensions fulfill $D_q^\mu = 1$, and the two inequalities are no longer good lower bounds. We like to point out that the decay of the center of the wave packet is only described by the correlation dimension $D_2^\mu$ of the LDOS until the integral in $C(t)$ converges \cite{JPhys.1995.Zhong}, i.e., as long as the exponent $\delta^\prime$, which describes the decay of the center of the wave packet, is smaller than 1. Hence, for $d$-dimensional systems the bound by Ketzmerick et al. should be replaced by $\beta \ge \delta^\prime / D_2$ in the absolutely continuous regime.
For the hypercubic tiling this bound is as good as for the Fibonacci chain. This bound should hold also in general because it solely makes use of the normalization condition of the wave packet. We have checked in Figure \ref{fig:bounds-beta} whether this relation is satisfied for the labyrinth tiling and found that this is a significantly better lower bound for $w > w_{\rm th}$.

\section{Conclusion}

We have studied the wave-packet dynamics for separable quasiperiodic tilings. In higher dimensions $\delta \to 1$ indicates the existence of an absolutely continuous part in the energy spectrum rather than the occurrence of ballistic transport. To understand the diffusive properties of a system it is also necessary to compute the width $d(t)$ of the wave packet and the return probability $P(t)$ because in three dimensions the scaling exponents $\beta$ hardly differ for the considered models although the scaling exponents $\delta^\prime$ are rather different for large coupling parameters $w$.
We also found good agreement between the scaling exponents $\delta^\prime$ and $\beta$ with the analytical expressions derived by an RG approach in the regime of strong quasiperiodic modulation. Further, the exponent $\delta^\prime$ can be used to define a better lower bound for the scaling exponent $\beta$ of the wave-packet width in the absolute continuous regime.

\end{document}